# What Are Chinese Talking about in Hot Weibos?

Yuan Li[1]    Haoyu Gao[2]    Mingmin Yang[2]    Wanqiu Guan[1]
Haixin Ma[3]    Weining Qian[3]    Zhigang Cao[2]    Xiaoguang Yang[1,2]

(1 College of Engineering and Information Technology, University of Chinese Academy of Sciences, 100049, Beijing, China)
(2 Academy of Mathematics and Systems Science, Chinese Academy of Sciences, 100190, Beijing, China)
(3 Institute of Massive Computing, East China Normal University, 200062, Shanghai, China)
(Emails: m_liyuan@163.com, gaosteveneric@gmail.com, yangmingmin2005@126.com, guanwanqiu10@mails.ucas.ac.cn, 51111500010@ecnu.cn, wnqian@sei.ecnu.edu.cn, zhigangcao@amss.ac.cn, xgyang@iss.ac.cn)

**Abstract:**    SinaWeibo is a Twitter-like social network service emerging in China in recent years. People can post weibos (microblogs) and communicate with others on it. Based on a dataset of 650 million weibos from August 2009 to January 2012 crawled from APIs of SinaWeibo, we study the hot ones that have been reposted for at least 1000 times. We find that hot weibos can be roughly classified into eight categories, i.e. Entertainment & Fashion, Hot Social Events, Leisure & Mood, Life & Health, Seeking for Help, Sales Promotion, Fengshui & Fortune and Deleted Weibos. In particular, Leisure & Mood and Hot Social Events account for almost 65% of all the hot weibos. This reflects very well the fundamental dual-structure of the current society of China: On the one hand, economy has made a great progress and quite a part of people are now living a relatively prosperous and fairly easy life. On the other hand, there still exist quite a lot of serious social problems, such as government corruptions and environmental pollutions. It is also shown that users' posting and reposting behaviors are greatly affected by their identity factors (gender, verification status, and regional location). For instance, (1) Two thirds of the hot weibos are created by male users. (2) Although verified users account for only 0.1% in SinaWeibo, 46.5% of the hot weibos are contributed by them. Very interestingly, 39.2% are written by SPA users (who behave quite abnormally in a way that what they post are almost always of the same type, say words of wisdom, "chicken-soup-soul" like sentences, and jokes etc.). This complements the previous finding of Yu et al. (2012), and may imply that SinaWeibo is in an "artificial inflation" not only on the reposting side but also on the posting side. A more or less pathetic fact is that only 14.4% of the hot weibos are created by grassroots (individual users that are neither SPA nor verified). (3) Users from different areas of China have distinct posting and reposting behaviors which usually reflect very their local cultures. Homophily is also examined for people's reposting behaviors.

## 1 Introduction

There has been a tremendous rise in the growth of online social networks service (SNS) all over the world in recent years. Online SNS today is playing an important role in people's daily lives. Also it has become a major platform, on which users can express their feelings and emotions, comment on their friends, play online games, join online community, and what's more, have their digital lives recorded [1].

Two worldwide popular SNS websites, Twitter and Facebook, have demonstrated their explosive growths and profound influences. While in China, SinaWeibo[1], emerging in August 2009, is also a typical SNS website and becoming more and more popular. It also has a length limit of 140 characters. Users can follow others, repost weibos to their homepages, and broadcast their own weibos to followers. If users want someone to read their messages, they can "@" his/her ID in the message to give a reminding. A prominent feature of SinaWeibo is that there are "VIP" accounts, which are called "verified users" (this mechanism is invented by SinaWeibo and followed by Twitter). Users who want to be verified could send an application and give a proof of their identities. If they get the official approval, there will be a sign of "V" after their names. There are also some differences between SinaWeibo and Twitter. For example, users on SinaWeibo can comment under the message, which will not be broadcasted to their followers. But on Twitter, users cannot comment without reposting.

---

* The research is supported by the 973 Program (2010CB731405) and NNSF of China (71101140).

[1] The web site: www.weibo.com

By August 2011, the number of registered users of SinaWeibo had reached 250 million, and the volume of weibos[2] per day had increased to 90 million, with almost 50,000 companies having opened brand accounts.[2] Facing such a huge amount of information, different users are interested in different weibos, and accordingly their behaviors vary greatly. For this whole new social world, there are obviously many interesting problems waiting for exploration. For instance, what kinds of topics are the hottest? What kinds of topics attract the most attention of people? What identity factors of users affect their posting and reposting behaviors and how? Do male users tend to repost more weibos created by males or the opposite? These are the main concerns of this paper.

To answer these questions, we collected 6025 original weibos (each has been reposted more than 1000 times and the total amount of reposted weibos is about 11 million). We found that these weibos can be roughly divided into eight categories, namely Entertainment & Fashion, Hot Social Events, Leisure & Mood, Life & Health, Seeking for Help, Sales Promotion, Fengshui & Fortune and Deleted Weibos. In particular, Sales Promotion and Seeking for Help topics are the hottest, being reposted on average by more than 2500 times, although they account for only 8% of the 6025 original weibos. The other two categories, Leisure & Mood and Hot Social Events, contribute almost 65% of the original weibos and more than 60% of the reposted weibos. This reflects very well the fundamental fact of today's China that on the one hand, economy has made a great progress and quite a part of people are now living a prosperous and fairly easy life, and on the other hand, there still exist quite a lot of serious social problems such as government corruptions and environmental pollutions.

As to the third question, we considered three identity factors of users, i.e. their genders, verification status and regional locations. Gender is doubtless a critical factor that might affect people's interests. Majority of the verified users are "elite" people in the real world. Many of them are public celebrities, and some others are organization accounts. Therefore, it is intuitive that verified users have more influence than unverified ones, and there might be a large difference between the behaviors of verified users and unverified ones. Regional location is also crucial because the development of China is nowadays quite unbalanced with respect to geographic location and this factor may influence their interests and behaviors greatly. We found that users' posting and reposting behaviors are significantly affected by these identity factors: (1) Two thirds of the hot weibos are created by male users. On the other hand, 51.1% of the reposted weibos are released by females. Combining the above two facts, we conclude that most of the hot weibos are created by male users while they become hot mainly driven by females. (2) Although verified users account for only 0.1% in SinaWeibo, 46.5% of the hot weibos in our dataset are contributed by them. What's more, 39.2% are written by very special users (who behave quite abnormally in a way that what they post are almost always of the same type, say words of wisdom, "chicken-soup-soul" like sentences, and jokes etc., and will be referred to as SPA users, see Section 4). Only 14.4% of the hot weibos are posted by grassroots (individual users that are neither SPA nor verified). This complements the previous finding that "a large percentage of trends in SinaWeibo are due to the continuous retweets of small amount of fraudulent accounts" (Yu et al, [10]). That is, SinaWeibo is in the "artificial inflation" not only in the reposting side but also on the posting side. (3) Users from different areas of China have distinct posting and reposting behaviors which usually reflect very well their local cultures. For each of the three factors, i.e. gender, verification status and regional location, homophily is observed for people's

---
[2] Weibos are the microblogs posted by users, like tweets in Twitter.

reposting behaviors, which answers the last question.

To end this section, we give a brief overview of related literature. With the births of Facebook (2004) and Twitter (2006), the studies of SNS are attracting more and more attention in recent years. Studies on Facebook can be found in [3] [4]. A seminal paper on Twitter is by Kwak et al. (2010) where they find interestingly that Twitter is more of a news media than a social network. They show that the majority (over 85%) of topics are headline news or persistent ones in the real world [5]. Wu et al. (2011) [6] focus on the question of "who says what to whom". They find that different people have quite different attention distributions on news topics, and demonstrate that significant homophily exists: celebrities listen to celebrities while bloggers listen to bloggers, etc.

Research on SinaWeibo began to emerge in 2011, two years after its birth. Chen et al. (2011)[7] and Gao et al. (2012)[8] make comprehensive comparisons between SinaWeibo and Twitter. Yu et al. (2011)[9] study the topic trends in SinaWeibo, and find that there is a vast difference between the contents shared in SinaWeibo and those in Twitter. In China, the trends are created almost entirely by reposted weibos of contents like jokes, images and videos, whereas on Twitter the trends tend to have more to do with current global events and news stories. In their later work (2012) [10] it is shown that reposted weibos (accounting for 65%) contribute a lot to creating trends and a large percentage of trends are due to the continuous reposts of a small amount of fraudulent accounts. Gao et al. (2012) [11] design and implement a real-time search system for a microblog system. Zhang et al. (2012)[12] propose an approach to generate the lurking users' profiles by its followees' activities and present a unified social context graph model to represent the lurking users' followees' activities. Song et al. (2012)[13] use machine learning tools to predict users' forwarding behavior. Chen et al. (2012)[14] investigate SinaWeibo by comparing the microblogging behaviors between verified users and unverified ones and study the social network evolutions of these two types of users. Chen et al. (2012)[15] find that SinaWeibo has a low reciprocity ratio, and users on Sina Weibo are more likely to follow people at higher or the same social levels. In addition to content and behavior studies, structure of the network of SinaWeibo is also investigated. For instance, Guo et al. (2011)[16] find, in SinaWeibo, that most links between users are one-way and there is a core network. For other famous SNS websites in China, they also attract some attention, e.g. TencentWeibo[17] and RenRen.com[18].

Homophily, a fundamental and pervasive phenomenon, refers to a tendency of individuals to interact more frequently with those similar to themselves, and limits people's social interactions in a way that has a big influence on the information they receive, the attitudes they form, and the relationships they construct. Homophily has been documented quite a lot for characteristics such as gender, race, ethnicity, age, class backgrounds, and educational level [19] [20] [21] [22]. Recently, Onnela et al. (2011)[23] conduct a research to evaluate the nature of the regional restriction on social group structures. By analyzing the mobile phone call and text messaging data from a European country, they argue that geography can be seen as a kind of constraint with respect to group formation. Feng Fu et al. (2012) constructed a model to show how natural selection can give rise to homophily when individuals engage in social interaction[24]. Kwak et al. (2010)[5] investigate homophily in two contexts: geographic location and popularity. They observe that users with followers 1000 or less are likely to be geographically close to their reciprocal-friends and also have similar number of followers.

## 2 Data description and preprocessing

The data we use are collected via APIs provided by SinaWeibo and the sampling method is the popular snowballing. We started from 32 prestigious users (11 public intellectuals and 21 academic scholars), obtaining their profiles, posted weibos, followers and followees. Then for each of the followers and followees, we obtained her/his the same set of information as above. This process was repeated three times, i.e. there are four layers of users. A total of 1.6 million users and 650 million original and reposted weibos (created during August 2009 and January 2012) were crawled. Then we extracted a collection of hot weibos that are reposted for 1000 times or more. This new set consists of 6025 original weibos (posted by 2356 users), forwarded for 11.08M times (by more than 11M users) in total (i.e. averagely 1839 per weibo). Among the 6025 original hot weibos, only two are posted in 2009, 229 in 2010, 5644 in 2011, and 150 in 2012 (only in January).

Our main preprocessing is that the weibos are classified into eight groups with respect to their topics (Table 1). This classification is done manually, hence the accuracy can be highly guaranteed. It is valuable to remark that, in reality and literature, there exist various classification approaches. For instance, in Sina ranking list[3], there are 12 categories (entertainment, sport, finance, game, travel, music, technology, auto-mobile, house, culture, life, fashion). This approach seems kind of both redundant and incomprehensive for us. The approach taken by Tencent weibo square[4], with five categories (information & sport, entertainment & video, life & fashion, culture & education and fresh point), however, is overly rough. In Twitter[5], there are no classification for tweets at all, but providing 30 categories for bloggers (e.g. music, sport, entertainment, funny, fashion, family, staff picks, charity, television and so on). In the research of Kleinberg et al. [25], most mentioned tweets in Twitter are also classified into eight categories of topics, i.e. Celebrity, Game, Idiom, Movie/TV, Music, Political, Sports and Technology. It can be seen that there are quite a few differences between their method and ours. A fairly authoritative report on online hot social events in the third quarter of 2011[26] divide these events into 11 categories (disaster, legal issues, public health, corruption, personal issues, social security, culture and education, livelihood, finance, international news and daily administration). But this classification applies only to online events. Our data, however, covers quite diversified topics, not restricted to events.

Table 1. Classification of hot weibos

| Label | Categories | Description |
|---|---|---|
| 1 | Entertainment & Fashion | fashion trends, entertainment videos, gossips about celebrities |
| 2 | Hot Social Events | natural disasters, public health, government corruptions, social security |
| 3 | Leisure & Mood | humors, literature and pictures, constellation, funny things, words of wisdom |
| 4 | Life & Health | common senses in life, healthcare, environmental protection |
| 5 | Seeking for Help | donations, searching for lost kids/old people |
| 6 | Sales Promotion | advertisements, network marketing |
| 7 | Fengshui & Fortune | these weibos promise that if you repost them, you will be blessed and can get good luck |
| 8 | Deleted | weibos deleted by administrators or users themselves, labeled as "This weibo has been deleted" |

For each original weibo, besides the user's gender, verification status and regional location, we also record whether it contains a URL, and whether a picture is included. Table 2 is an illustration of our data set, where we list the top 20 hottest weibos.

Table 2. Top 20 hottest weibos.

| Weibo ID | User ID | Gender | Verification | Region | URL | Picture | Category | Reposted times |
|---|---|---|---|---|---|---|---|---|
| 3339358618814023 | 1969451625 | Male | √ | Guangdong | × | √ | 2 | 35066 |
| 3371677214124537 | 1821898647 | Female | √ | Beijing | √ | √ | 6 | 33154 |

---

[3] http://data.weibo.com/top?topnav=1&wvr=4
[4] http://c.t.qq.com/k/default?pgv_ref=web.wide.page.nav.text
[5] https://twitter.com/#!/who_to_follow/interests

| | | | | | | | | |
|---|---|---|---|---|---|---|---|---|
| 5618714962225071290 | 1864419143 | Female | √ | Beijing | √ | √ | 6 | 30339 |
| 5598249344277385949 | 1764452651 | Male | √ | Shanghai | × | √ | 5 | 29610 |
| 2011003182311915 | 1152759471 | Male | × | Beijing | × | √ | 8 | 23363 |
| 3369183448905874 | 1649155730 | Male | √ | Guangdong | × | √ | 6 | 21475 |
| 3385851792713190 | 1677969704 | Male | √ | Beijing | √ | √ | 6 | 19757 |
| 221110325332552883 | 1686326292 | Female | √ | HongKong | × | √ | 5 | 18321 |
| 3369783744414548 | 1911718510 | Male | √ | Oversea | × | √ | 5 | 18231 |
| 5620644665262856023 | 1404411345 | Male | √ | Shanghai | √ | √ | 3 | 18109 |
| 3339094088416508 | 1991163697 | Male | × | Zhejiang | × | √ | 5 | 17974 |
| 3352116729032709 | 1864419143 | Female | √ | Beijing | √ | √ | 6 | 16904 |
| 3391244082844889 | 1641532820 | Male | √ | Guangdong | √ | √ | 5 | 15799 |
| 5619360762097330512 | 1189591617 | Male | √ | Other | √ | √ | 3 | 15575 |
| 5603664396091205542 | 1692544657 | Male | √ | Taiwan | × | √ | 3 | 15231 |
| 5626163182999460000 | 2089106454 | Male | √ | Other | × | √ | 2 | 14318 |
| 5614379729777485798 | 1638781994 | Male | √ | Beijing | × | √ | 1 | 13445 |
| 5615734448357004557 | 1891845373 | Male | √ | Shanghai | √ | √ | 6 | 13224 |
| 5607334484304271198 | 1525396581 | Male | √ | Beijing | × | √ | 2 | 12935 |
| 5613056931220737005 | 1650507560 | Male | √ | Guangdong | × | √ | 5 | 12686 |

## 3 Overall analysis

In this section, we shall give an overall analysis and try to answer which topics are the hottest. To measure the hotness of a single weibo, doubtlessly we should use the number of reposted times. To measure the hotness of a category of topics, however, there can be more than one natural approaches. As displayed in Table 3, we took four indices, i.e. the total number of hot original weibos (as well as its proportion in 6025), the average number of reposted times, the maximum number of reposted times, and the total number of reposted times (as well as its proportion in 11M).

More than 60% of all the hot weibos fall into the category of Leisure & Mood (42.6%) and Hot Social Events (21.9%). This reflects very well the fundamental fact of today's China that on the one hand, economy has been greatly advanced and quite a part of people are now living a prosperous and fairly easy life. This is especially so for SinaWeibo users, who use computers/smart phones and Internet frequently at work and home. They seek for fun and talk about "merry" things in SinaWeibo. This is the main aim they register Weibo and also one of the basic functions of SinaWeibo. On the other hand, there still exist quite a lot of "serious" topics, for example, government corruptions, social security problems, public health issues and natural disasters. Chinese people really need a platform to express their emotions and opinions and make their voice heard. Internet serves this platform irreplaceably, and in particular, SinaWeibo, due to its strong interaction and social feature, is doing a better job. Facts tell us that this works, although censorships by no means disappear. For example, quite a few of government corruptions are disclosed and widely spread via SinaWeibo, and they were eventually handled rather properly. This leads to the phenomenon of "cyber politics" and promotes significantly the progress of society.

Entertainment and fashion are both huge industries in the real world. Being different with most other industries, their successes rely largely on people's attention. How successful a film star is can be roughly measured by and is mainly determined by how many people are talking about him/her. Therefore many entertainment companies take good use of this new media of SinaWeibo to "sell" their members. On the other side, lots of ordinary people are very interested in the private lives of the stars and enjoy gossiping them. In a word, they do "buy" this kind of advertisement. The above two forces make it that more than 10% of the hot weibos are about Entertainment & Fashion.

For companies in other industries, some of them are also trying to take advantage of SinaWeibo to do marketing. This is the so-called "Weibo Marketing". This phenomenon is reflected in the category of Sales Promotion. Although hot weibos in this category amount to only 3.3%, their average reposted time is the highest. There are still three other categories whose total numbers of original hot weibos are small but the average reposted times are high, namely Fengshui & Fortune, Seeking for Help, and Deleted. The original weibos about Fengshui & Fortune are quite abnormal. They promise that you will be blessed or get good luck if you repost it (and very rarely, it curses that you will be doomed or get bad luck if you neglect it). A large number of people just do not want to ignore the good promise (or curse) and tend to repost as required (or forced). This is why these weibos are reposted so many times. The topics about Seeking for Help, which are mostly looking for lost old people and kids, reflect some of the most serious social problems China is facing today, say fast-urbanization, gender imbalance of population, and birth-control policy. In addition, it also shows that Weibo promotes the social progress, and the majority of people are full of love and willing to help. Two other categories of topics, Life & Health and Deleted are also hot. The former reflects one of the key functions of the Internet as a platform of spreading information and knowledge. Weibos in the latter category are mostly politically sensitive. These topics are currently not allowed to discuss publicly by the authorities. In SinaWeibo, these sensitive topics are simply deleted. The fact that these topics are spread widely before deletion indicates that there are still many people that are interested and would like to repost them.

Let's now explore the distribution of reposted weibos for each topic. Considering that the weibos that are reposted less than 1000 times are not included in our data set, simply putting the raw data into the log-log space will result in a "fat-tail" which is quite uninteresting. So we bin the reposted times with interval 100, i.e. for each positive integer $i$, frequencies of reposted times falling into the interval of [1000+100($i$-1), 1000+100$i$) are summed up as the new frequency of 1000+($i$-1)100. The distribution after this treatment is illustrated in the left part of Figure 1. We can see that the new points fit Power Law nicely. Moreover, for each category, its empirical CDF (cumulative distribution function) is also displayed in the right part of Figure 1. It is easy to observe that these CDFs can be classified into two groups. The upper group of categories, each of which has been averagely reposted less than 2000 times, consists of Entertainment & Fashion, Leisure & Mood and Life & Health, while the lower group of categories, each of which has been reposted more than 2000 time on average, includes the rest five. This is consistent with the basic statistics in Table 3.

Table 3. Basic statistics of hot weibos

| Classification | Entertainment & Fashion | Hot Social Events | Leisure & Mood | Life & Health | Seeking for Help | Sales Promotion | Fengshui & Fortune | Deleted |
|---|---|---|---|---|---|---|---|---|
| Original weibos | 809 (13.43%) | 1320 (21.91%) | 2569 (42.64%) | 448 (7.44%) | 290 (4.81%) | 201 (3.34%) | 212 (3.52%) | 176 (2.92%) |
| Reposted (maximum) | 13445 | 35066 | 18109 | 7462 | 29610 | 33154 | 10204 | 23363 |
| Reposted (average) | 1721 | 2044 | 1606 | 1586 | 2572 | 2611 | 2172 | 2312 |
| Reposted (total) | 1392989 (12.59%) | 2698553 (24.38%) | 4127854 (37.29%) | 710848 (6.42%) | 745954 (6.74%) | 524937 (4.74%) | 460573 (4.16%) | 406961 (3.68%) |

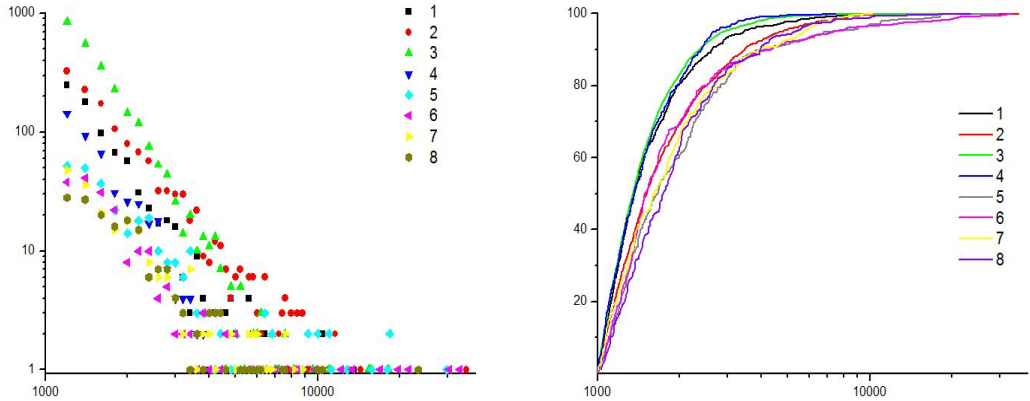

**Figure 1. Distributions and CDFs of reposted times**

To end this section, we give some multi-media analysis. Due to that users on SinaWeibo are allowed to insert pictures and URL in every message, we look into the proportions of them in our data. Noticeably, a large part (78%) of the original hot weibos contains pictures. We guess that there are mainly two reasons for this. First, as the western proverb goes, "A picture says more than a thousand words", and hence weibos with pictures usually contain much more information than those without pictures. Second, as accepted widely in China's Netizens, "Pictures don't lie" (有图有真相), and hence weibos with pictures are much more trustworthy. Only 23% original hot weibos have URLs. This point is a little difference from Twitter. Twitter users post URLs in only 17.6% of the tweets on trending topics [9].

## 4 Identity factors

To proceed with analysis of this section, we need several new terms. We call users who have posted at least one original weibos in our data "creative users". We find that 69% of the creative users are registered as IA (Individual Accounts) contributing 44% weibos, and 15% as OA (Organization Accounts) posting 17% weibos. The rest 16% users who release 39% weibos can be hardly called OA users or IA ones. In the rest of this paper, we call them SPA (Special Purpose Accounts). On the one hand, these users are not registered as organizations, and it is very likely that there are no organizations in the real world that correspond to these accounts. On the other hand, they behave quite differently with normal individual users: what they post are almost always of the same type, say words of wisdom, "chicken-soup-soul" like sentences, and jokes etc. And there is no actual interaction between them and their followers. In a word, their behavior is quite stylized, and we guess that the meaning of their existence is simply to attract followers. Some of them may be spammers, but it is not an easy task to confirm this.

The basic statistics for the three types of accounts are shown in Table 4. Our definition of SPA can be perfectly justified from this table. On the one hand, the verified ratio in SPA is extremely low, while this ratio in OA is extremely high, so SPA users are rather distinct from OA ones. On the other hand, the male ratio in SPA is much lower than that in IA.

**Table 4. Statistics of accounts in original hot weibos**

| Users | Male | Female | Verified | Unverified | Total |
|---|---|---|---|---|---|
| IA | 1157 (71.33%) | 465 (28.67%) | 942 (58.08%) | 680 (41.92%) | 1622 (68.84%) |
| OA | 235 (66.38%) | 119 (33.62%) | 351 (99.15%) | 3 (0.85%) | 354 (15.03%) |
| SPA | 176 (46.32%) | 204 (53.68%) | 9 (2.37%) | 371 (97.63%) | 380 (16.13%) |
| Total | 1568 (66.55%) | 788 (33.45%) | 1302 (55.26%) | 1054 (44.74%) | 2356 (100%) |

Next, we shall illustrate that identity factors of users, namely gender, verification status and regional location, affect their posting and reposting behaviors significantly.

### 4.1 Gender effect

Above all, although gender attributes of SPA and OA users are all filled out (as required by SinaWeibo when registered), we think that these information is noisy for us to analyze gender effect on users' posting and reposting behaviors, so we deleted the original hot weibos posted by them (however, for technical reasons, we did not do this treatment for reposted weibos).

The DCCI 2011 Report shows that in SinaWeibo, male users are in majority, accounting for 57.4%. But in our data, the proportion of male creative users in original hot weibos, 71.3%, is much higher. It shows that weibos posted by male users are more likely to attract others' attention and to become hot. Moreover, through our statistics, these male users contribute 72.8% of the original hot weibos.

For each of the eight categories, we further count the numbers of weibos posted by males and females, respectively. The primary part of Figure 2 shows that for each category, the number of weibos posted by males is much larger than that by females. Considering that males are more than females in the whole world of SinaWeibo, we normalize these numbers with two methods. In the first method, we calculate the proportions of each gender in every category (the upper embedded bar graph in Figure 2). In the second one, we compute male's weibo distribution in all categories as well as the female's (the lower embedded bar graph in Figure 2).

From the upper embedded bar graph, we find that the proportion of males for each category is significantly larger than the baseline, 0.57 (DCCI, 2011). It illustrates that male users are more likely to create original weibos for any topic and they have more creativity than female ones when they post weibos. Moreover, the proportion gap between male and female is the largest in Sales Promotion.

From the bottom one, it can be observed that both male and female users write quite a lot weibos on Leisure & Mood (40%). This is plausible, because the main function of SinaWeibo is to serve as a platform for people to express their daily feelings. However, male users distribute more than one third of their effort on Hot Social Events and females allocate less than one fourth. In addition, male users are more willing to post these topics that mainly Sales Promotion than female (although both proportions are rather small). Female users write relatively more on weibos of any other categories, though female users post absolutely less weibos than males.

In order to explore the gender effect on reposting behaviors, for each category of hot weibos, we count the total number of weibos reposted by male users and that by females (Figure 3). From it, we can find that except for Hot Social Events, Sales Promotion and Deleted, the number of weibos reposted by female users is significantly greater than by males. As in Figure 2, two kinds of normalizations are presented. It can be found from the upper embedded graph that only in Hot Social Events, male's proportion is a little over the baseline 57%. Meanwhile, the average of female's proportions, 51.1%, although only slightly bigger than 50%, is significantly larger than 43% (the female ratio in SinaWeibo). From the bottom embedded graph, we can see that males and females both prefer paying much attention to Entertainment & Fashion, Hot Social Events and Leisure & Mood, especially Leisure & Mood. This result is the same with the posting side. But males are more likely to pay attention to serious and social topics, while females tend to care more about topics involving relaxing and personal contents. For Seeking for Help, males and females pay almost the same proportion of their attentions. This indicates sympathy, as a human nature, exists universally and evenly in males and females.

As a brief conclusion of this subsection, most of the hot weibos are created by male users while they become hot mainly driven by females.

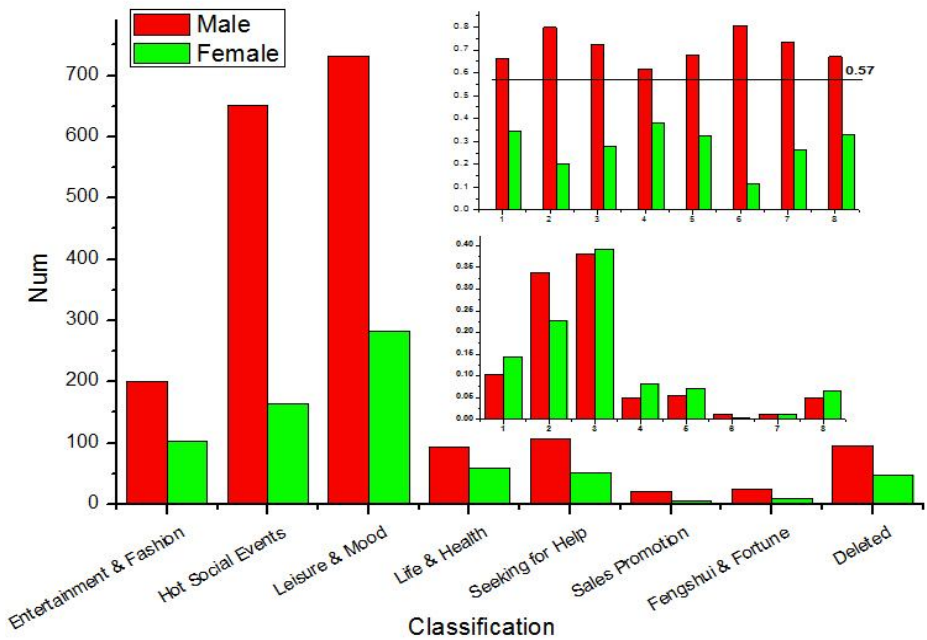

**Figure 2. Numbers of weibos posted by males\females**

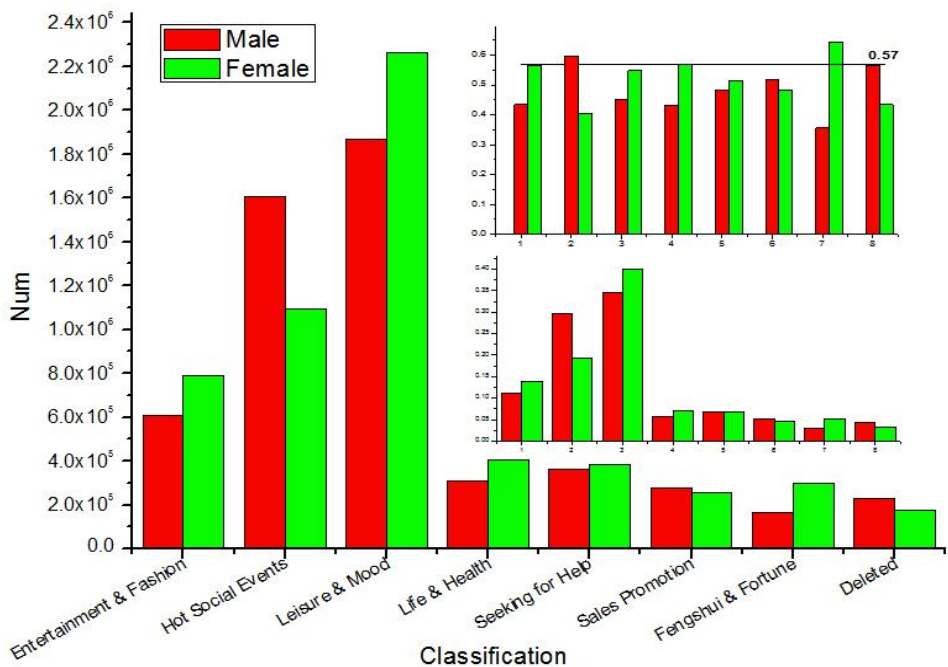

**Figure 3. Numbers of weibos reposted by males\females**

**4.2 Verification effect**

Until 2011, the total number of verified users in SinaWeibo has grown to 300 thousand [12], while the total users are over 250 million [2]. That is to say, verified users only account for about 0.1% in SinaWeibo. But from Table 4, in our data, they occupy 12.3% of all the users (and 55.3% of all the creative users!). This is an astonishing gap! It is thus clear that verified users are much more likely to appear in hot weibos. This difference also shows that compared with unverified ones, verified users tend to catch more attention. On the other hand, among all the hot weibos in our data, 46.5% are contributed by verified users. This is even more striking than the finding of Bill Heil et al. (2009) which shows that "*the top 10% of prolific Twitter users account for over 90% of tweets*" [27]. It should not be neglected that there are still a great many creative users who are unverified. Due to the analysis below, there may be another reason, that is, lots of hot weibos in our data (39.2%) are created by SPA users, who are usually unverified. However, if deleted the weibos posted by SPA users, 76.2% of remaining weibos are released by verified users.

Let us now concentrate more carefully on the "posting" side. For each category of original hot weibos, we use the same method as in the previous subsection and get Figure 4. For Entertainment & Fashion, Leisure & Mood, Life & Health, and Fengshui & Fortune, weibos posted by unverified users are more than those by verified ones. Oppositely, for Deleted and Hot Social Events weibos, verified users are more willing to post them.

In the top embedded bar graph of Figure 4, not surprisingly, for each category, the proportion of weibos posted by verified users are much higher than the baseline of 0.1%. However, two interesting observations are still worth of pointing out. To be specific, (i) for both categories of Hot Social Events and Sales Promotion, verified users play even dominating roles (accounting for more than 70%). This is quite understandable, because verified users are generally elites and hence they are more interested in public affairs. As to why the proportion of verified users for Sales Promotion is so strikingly high (90%), it is sufficient to note that these verified users are mostly OA. (ii) Compared with other categories, for Fengshui & Fortune, the proportion of verified users is relatively low (although also much larger than 0.1%).

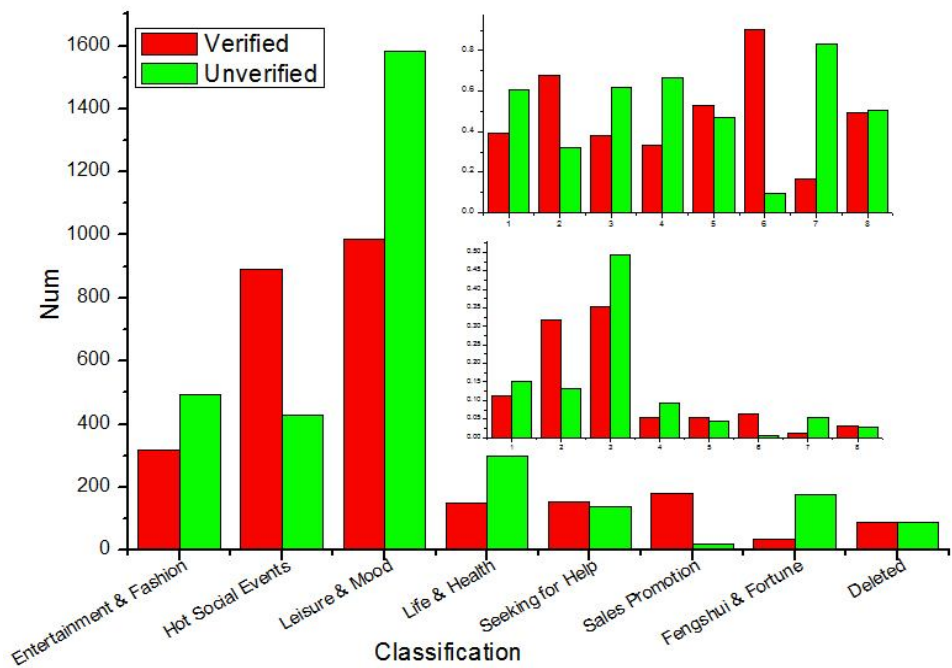

**Figure 4. Numbers of original weibos posted by verified\unverified users**

Now, let's turn our attention to the reposting side. As before, we get Figure 5 which describes the numbers of weibos reposted by verified users and unverified users and draw two embedded bar graphs. From Figure 5, we can see that weibos reposted by unverified users account for the vast majority for all categories. Quite unexpectedly, proportions of verified users in all the eight categories are around 12.3% with very little fluctuation, except for Sales Promotion. We guess that this category is special because verified users have very little interest to get rewarded by reposting promotion advertisements. Tiny promotion rewarding has little temptation to the elite people.

We conclude this subsection to note that verified users are much more influential in the online world, as in the real world. They shoulder more social responsibility and show more interests in hot social events and politically sensitive affairs. They are also more willing to give a hand to those in trouble via posting and reposting the "Seeking for Help" weibos, because they know they own large number of followers and thus can spread the information very quickly.

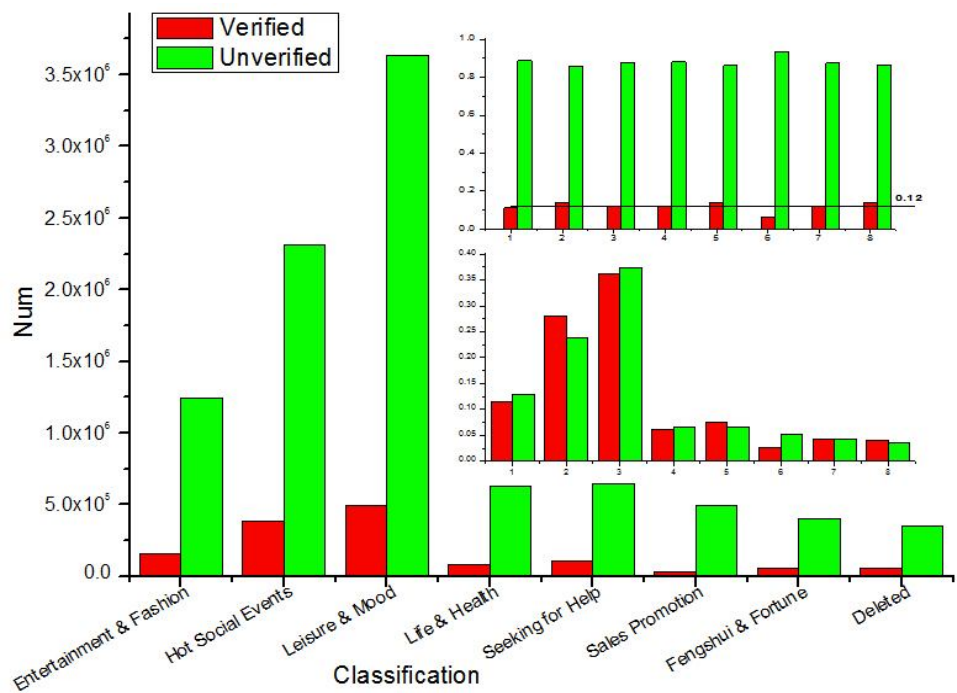

**Figure 5. Numbers of weibos reposted by verified\unverified users**

## 4.3 Regional effect

Regional analysis is very necessary, because China is a huge country and different regions usually have quite different cultures and different living styles. So it is natural to ask whether these differences are reflected in SinaWeibo.

To answer this question, for each province and each municipality directly under the central government (say Beijing, Shanghai), we count the total number of original hot weibos. It is interesting to notice that quite a few creative users are registered as from Hong Kong, Taiwan, and overseas. Although we cannot guarantee that these information are all valid (for instance, many people might register their location as overseas just for fun), we do not do any further treatment (as for SPA) because this is technically impossible for us. Another point that should be noted is that about 3.9% creative users did not report their regional information, and we label them "Other" as they intentionally chose when registered.

It can be seen clearly from Figure 6 that Beijing, Shanghai and Guangdong are the top three regions in our data. And regions that are ranked in the front are all economically developed areas. One reason is that the total user numbers of these provinces in SinaWeibo are much larger than other places. Another possible reason is that for our data, users from these regions are more likely to create and repost hot weibos.

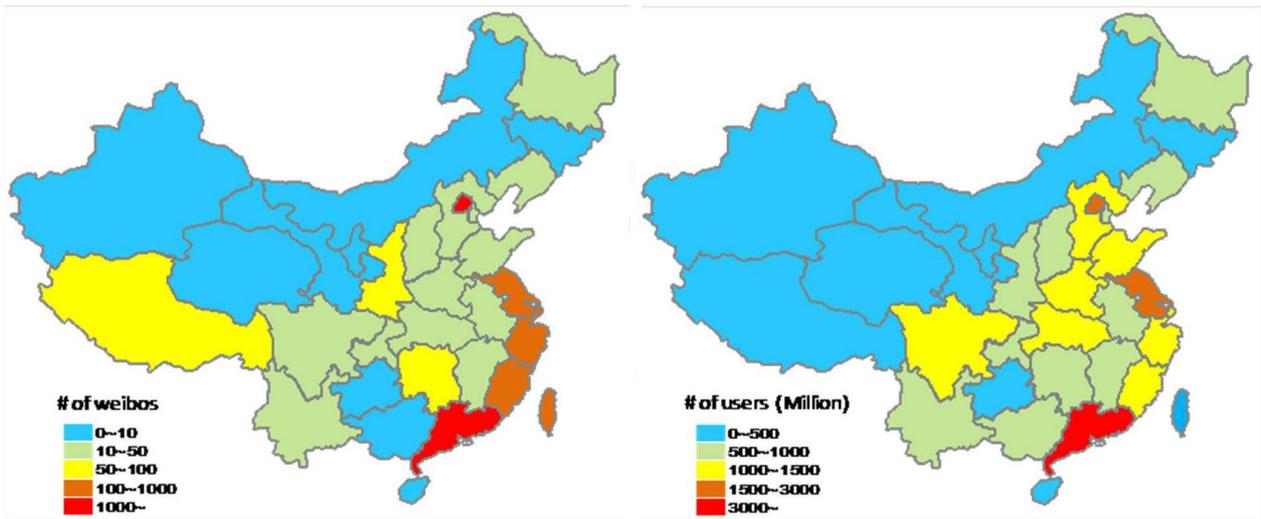

**Figure 6. Regional distributions of hot weibos (left) and users (right).**
(The right map is extracted from [28])

To do more detailed analysis about the distribution of regions for different categories of original hot weibos, we use the same way as before to count the number, proportion and distribution of weibos posted by users in each region. All the results are shown in Figure 7. It can be observed that for each category, a majority of hot weibos are posted by Beijing, Shanghai and Guangdong users. We find that on Entertainment & Fashion and Leisure & Mood, users from HongKong and Taiwan attract significantly more attention. One reason might be that these users are mostly movie stars and singers, and they need to advertise their new movies or songs via SinaWeibo. As to Seeking for Help, the majority of users are from Zhejiang and Jiangsu, two developed areas. On the one hand, due to fairly high level education, whether both in the cities and the countryside, people are more willing easily to seek solutions through innovative new ways, such as the Internet. On the other hand, the family planning policy has been strictly implemented well in these areas. Even if in the countryside, most parents have only one, at most or two, children. So that the parents pay more attention to children whenever kids are missing, most families will try their best to find them back, using all methods they can image, including the internet of course. At the same time, weibos about Sales Promotions in the two provinces are also rather outstanding, because they have numerous small companies. Very interestingly, more than a half of hot weibos on Fengshui & Fortune are written by Guangdong users. This is nicely consistent with their local culture.

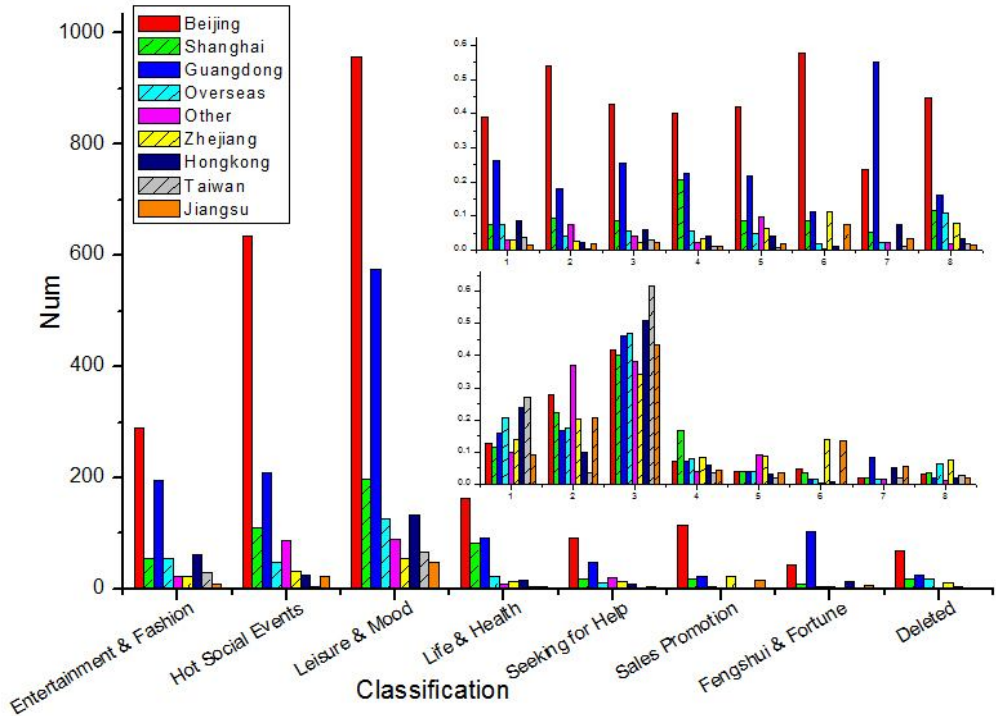

**Figure 7. Numbers of weibos posted by users from different regions**

## 5 Homophily

For SinaWeibo, it is obviously meaningful to see whether homophily exists. Specially, we investigate the same three factors studied in the previous section, and record these results in Table 5.

For each original hot weibo, we count the numbers of corresponding reposting weibos from males and females, respectively, and get the female ratio. We say that gender homophily exists for this weibo either if (i) it was created by a female user and its female ratio in all the reposting users for this weibo is larger than that in the whole world of SinaWeibo (43%), or (ii) it was written by a male and its male ratio in the reposting users is larger than 57%. For each category, we calculate the proportion of original weibos possessing gender homophily.

From Table 5, we can find that for all categories, gender homophily does exist, which means that users are more likely to repost weibos from those written by the same gender. This is especially true for Hot Social Events and Deleted, with nearly 70 percent weibos satisfying homophily. While for Seeking for Help, its homophily index is the smallest. This is plausible because people usually do not care whether a weibo is posted by a male or female when they consider to repost it to help someone in trouble.

For verification status, we also do the same treatment as for gender. Overall, the average ratio of verified users who repost hot weibos created by verified users, 13.5%, is larger than the base ratio of 12.3%. This ratio for weibos posted by unverified users is 11.0%, smaller than 12.3%. This means that verified users, in general, are (relatively) more likely to pay attention to verified ones, and vice versa. Similar to the factor of gender, seven out of eight categories have more than half of their weibos satisfy verification homophily. For Sales Promotion, the ratio is surprisingly low, this is because most of these weibos are created by OA users, who are almost all verified, and these verified users are usually not willing to repost other sales promotion weibos.

For an original weibo, say it is posted by a guy from province A, we calculate the proportions of weibos that are reposted by users from the same province, and those from all the other provinces. If (i) the ratio of province A is the highest of all, and (ii) it is larger than the ratio of province A users among all the users, we say that location homophily exists for this weibo. Note that this definition is stronger than the one we used for the factors of gender and verification status. For each category, we compute the ratio of original hot weibos where location homophily exists. As shown in Table 5, location homophily is significant for most of the categories. For Deleted (0.78), Seeking for Help (0.67) and Hot Social Events (0.68), they all have higher location homophily. These results are understandable, because the impacts of hot social events are usually restricted to the region where they happen, people are more willing to help those in trouble that are from their hometown, and deleted weibos are generally the ones that are politically sensitive, and particularly, regional politics related, say about violently forced relocations. If we use the public data released by Yiguan[29] as our users' location distribution benchmark, the results will be even more significant. What's more, if we relax the definition of homophily by dropping condition (i), then the above results are even more valid.

Table 5. Homophily ratios

| Classification | Entertainment & Fashion | Hot Social Events | Leisure & Mood | Life & Health | Seeking for Help | Sales Promotion | Fengshui & Fortune | Deleted |
|---|---|---|---|---|---|---|---|---|
| Gender homophily index | 0.551 | 0.708 | 0.617 | 0.643 | 0.545 | 0.617 | 0.557 | 0.716 |
| Verification homophily index | 0.606 | 0.606 | 0.671 | 0,723 | 0.583 | 0.149 | 0.642 | 0.585 |
| Region homophily index | 0.576 | 0.678 | 0.565 | 0.549 | 0.668 | 0.447 | 0.419 | 0.778 |

## 6 Conclusions

In this paper, based on a fairly large amount of data, we have given our answer to the question of what are Chinese talking about in hot weibos, and demonstrated several other interesting facts. We found that Leisure & Mood and Hot Social Events account for almost 65% of all the hot weibos, which reflects very well the fundamental dual-structure of the current society of China. It is also shown that users' posting and reposting behaviors are greatly affected by their identity factors (gender, verification status, and regional location). Homophily is also examined for people's reposting behaviors. This paper is only the first attempt towards the topic of hot weibos, and many other interesting problems, e.g., the distributions of reposting depths and reposting widths, properties of "tipping points" and diffusion processes of hot weibos, etc., are still widely open and waiting to be explored.